\documentstyle[epsfig]{aipproc}

\begin{document}
\title{Are The Four Gamma-Ray Bursts of 1996 October 27-29\\
Due to Repetition of a Single Source?}

\author{Carlo Graziani, Donald Q. Lamb and Jean M. Quashnock}
\address{Department of Astronomy and Astrophysics, University of
Chicago}

\maketitle

\begin{abstract}
BATSE, Ulysses, and TGRS and KONUS on WIND detected four gamma-ray
events within 1.8 days during 1996 October 27-29, consistent with
coming from the same location on the sky.  We assess the evidence that
these events may be due to a series of bursts from a single source by
calculating the probability that such a clustering in position and in
time of occurrence might happen by chance.  The calculation of this
probability is afflicted by the usual problem of a posteriori
statistics.  We introduce a clustering statistic, which is formed from
the "minimum circle radius" (i.e. the radius of the smallest circle
that just encloses the positions of all the events) and the minimum
time lapse (i.e. the time elapsed between the first and last event). 
We also introduce a second clustering statistic, which is formed from
the "cluster likelihood function" and the minimum time lapse. We show
that the use of these statistics largely eliminates the "a posteriori"
nature of the problem.  The two statistics yield significances of the
clustering of $3.3\times 10^{-4}$ and $3.1\times 10^{-5}$,
respectively, if we interpret the four events as four bursts, whereas
the clustering is not significant if we interpret the four events as only
three bursts. However, in the latter case one of the bursts is the
longest ever observed by BATSE.
\end{abstract}

\section*{Introduction}

The question of whether gamma-ray burst sources repeat on timescales of
a year or less has occasioned much debate since the public release of 
the first BATSE (1B) catalog \cite{fishman94}.  Information concerning
such repetition would prove enormously constraining on gamma-ray burst
models.  Most models of burst sources at cosmological distances cannot
plausibly repeat on such timescales, since in these models the burst
destroys the source.  Models of burst sources located in a galactic
halo, on the other hand, are expected to repeat on such timescales on
energetic grounds \cite{lbc95}.  Thus, confirmation of such
repetitions would create many difficulties for cosmological models,
while ruling them out would seriously constrain galactic halo ones.

There have been several claims and counterclaims made in the literature
on the subject.  Global evidence for repetition was found in the 1B
catalog but not confirmed in the 2B and 3B catalogs; individual clusters
suggestive of repetition were also found in the 1B catalog, although
their statistical significance was not impressive.  For details, see the
discussion in \cite{lamb96}.

\begin{figure}
\centerline{\epsfig{file=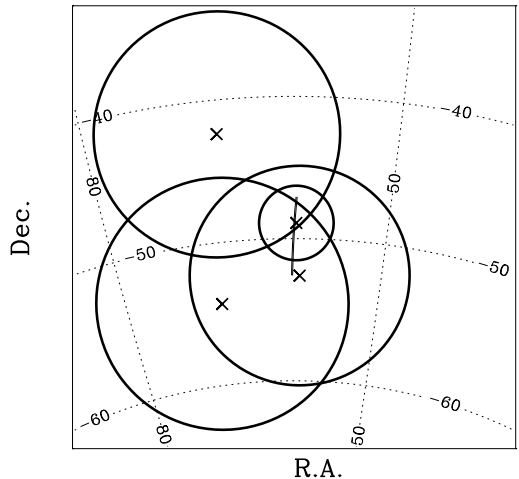,scale=0.58}}
\caption[fooba]{Sky locations of the four events.  The crosses
represent the BATSE locations.  The errors circle radii are computed
using the best-fit model for the BATSE location systematic error from
Graziani \& Lamb (\cite{gl96}).  The IPN location is also shown.}
\label{skyplot}
\end{figure}

This, then, was the situation on 1996 October 27, when the first of a
series of four BATSE GRB triggers occurred.  By the time of the fourth
trigger, 1.8 days later, the four burst triggers were conspicuous for
having occurred with a relatively short time and for having locations
consistent with a single joint source on the sky \cite{vc97}.  The
locations are shown in Fig. \ref{skyplot}. The positional and temporal
coincidence of these events is striking, and provokes the question of
whether it is probable that they were produced by a single, repeating
gamma-ray burst source.  We address precisely this question in what
follows.

%
%
%

\section*{The Data}

The four burst triggers (BATSE trigger numbers 5646, 5647, 5648, and
5649) occured in two pairs.  The first pair occurred about 18 minutes
apart on October 27, while the second pair occurred 11.2 minutes apart
1.8 days later, on October 29.  The last event was considerably more
intense and fluent than the first three, and was also detected by
KONUS-WIND and by ULYSSES, resulting in the truncated IPN annulus shown
in Fig. \ref{skyplot}.

While it is not unnatural to regard these four triggers as four
gamma-ray bursts, the BATSE team has raised the possibility that the
last two triggers may in fact correspond to one, extremely long
burst, in which case there would only be three bursts \cite{vc97}.
This possibility is difficult to assess, as there is no
model-independent way of making the distinction between a gamma-ray
burst trigger and a gamma-ray burst.  At a minimum, the claim that
triggers 5648 and 5649 are due to the same gamma-ray burst is open
to debate. Certainly, the triggers were initially classified as
independent bursts.  There does not appear to have been any emission in
the approximately 11 minutes between the end of the third event and the
onset of the fourth, and if the two events were part of the same burst,
it would be the longest burst observed to date by BATSE.

Nonetheless, the possibility exists that there were in fact only three
bursts observed during the period in question, and we will consider
this possibility in our analysis.  In the following, when we do
consider the case of only three events, we will do so by dropping the
third event (trigger number 5648), since trigger number 5649 was much
more fluent, so that its location is much more accurately determined.

\section*{Analysis}

We adopt the frequentist statistical methodology of hypothesis testing. 
The null hypothesis under scrutiny is that the four (or three) bursts
in question all had distinct sources.  It is necessary then to identify
a test statistic that allows us to ask some form of the question
``assuming the null hypothesis, what is the probability that the observed
position-time clustering should have occurred?''

The chosen test statistic should certainly be sensitive to the kind of
clustering of interest.  However, there exists a danger that the
statistic, being chosen \emph{a posteriori}, will be unfairly selected
to confer an excessive significance upon this particular data set. As
an example of this danger, consider the following calculation.  The
four BATSE point locations may be inscribed in a circle of radius
$6.1^\circ$, and their times of occurrence in a time window of duration
1.8 days.  Using a rate of BATSE burst triggers of $0.8$ day$^{-1}$,
this works out to an expected number of events in our
``time-solid-angle'' window of $4.1\times 10^{-3}$. The resulting
Poisson probability for seeing four events is $1.1\times 10^{-11}$. 
Multiplying by the number of such ``time-solid-angle'' windows in the
BATSE catalog up to 1996 October 29, we obtain the apparently
impressive significance of $6.9\times 10^{-6}$.

The problem with the above calculation is that it would produce greatly
different significances for different choices of the
``time-solid-angle'' window.  On the one hand, there is no unique way
to pick the window for this kind of calculation. On the other hand, the
choice made above is the one that (unfairly) maximizes the ostensible
significance of the result for the particular data that was observed.  

Guided by the above example, we have sought out time-solid-angle
clustering statistics that are not arbitrary, in the sense that they do
not contain parameters which may be adjusted independently of the data
to ``optimize'' the significance (such as the radius and duration of
the window in the example).  We have identified two suitable test
statistics:  the minimum circle radius and the cluster likelihood.

The minimum circle radius approach is a twist on the previous example.
Instead of calculating the probability that a 1.8~day-6.1$^\circ$
``time-solid-angle'' window in the BATSE catalog should contain four or
more bursts, we calculate the probability that there should be four
bursts in the BATSE catalog characterized by a \emph{minimum
circle} of radius 6.1$^\circ$ or less, and a first-to-last duration of
1.8~days or less.  By minimum circle, we mean the unique smallest
circle that just includes all of the events.  By this modification, we
eliminate the arbitrariness in the choice of the window parameters,
which are set by the data itself.  The probability may no longer be
calculated by the naive Poisson method described in the example. It is
necessary to calculate the distribution function for minimum circle
radii for Poisson point processes on the sphere.

We have calculated this distribution function.  Suppose the Poisson
process has a rate $R$, so that in a time $\tau$ it averages $R\tau$
events.  The probability that it should produce $n$ events inscribed by
a minimum circle of radius $\theta_0$ or less is given by
$Q_n(\theta_0;R\tau)$, where
\begin{equation}
Q_n(\theta_0;R\tau)=e^{-\bar{n}}\,\Bigg\{
\frac{R\tau\,n\,\bar{n}^{n-1}}{(n-1)!}
+ [R\tau\,n - (n-1)(n+2)]\,\sum_{l=n}^\infty\frac{\bar{n}^l}{l!}
\Bigg\},
\label{qn}
\end{equation}
where $\bar{n}\equiv R\tau(1-\cos\theta_0)/2$ is the expected number of
events in the circle.

We actually need the $F_n(\theta_0,\tau_0)$, the probability that $n$
events should inscribe themselves in a minimum circle of radius
$\theta_0$ or less \emph{and} should have a first-to-last duration
$\tau_0$ or less.  We have rigorously derived the expression for $F$.
Assuming that $\tau_0\ll T$, where $T$ is the lifetime of the
experiment, we find 
\begin{equation}
F_n(\theta_0,\tau_0) \approx 
Q_n(\theta_0;R\tau_0) \times \frac{T}{\tau_0} \times n.
\label{mct}
\end{equation}

Applying Eq. (\ref{mct}) to the data, we find that assuming there are
four events in the set, the probability is $3.2\times 10^{-4}$, while
assuming three events, the probability is $1.2\times 10^{-1}$.  We have
confirmed these results by Monte Carlo simulations --- $3.3\times
10^{-4}$ and $1.1\times 10^{-1}$, respectively.

We have as yet made no use of the information borne by the BATSE error
circles.  They may be introduced by means of the cluster likelihood, 
which was introduced in a Bayesian framework by Graziani \& Lamb
\cite{gl96b}.  Briefly, it is the probability that $n$ events should
have their measured locations assuming they have a common source,
integrated over their unknown source position.  In the Gaussian error
approximation, the expression for the cluster likelihood is
\begin{equation}
{\mathcal{L}}=\frac{\Gamma^2/2}{\prod_{i=1}^n\left({\sigma_i}^2/2\right)} \times
\exp\left[-\frac{1}{2} \sum_{i=1}^n
\frac{(\vec{x}_i-\vec{z})^2}{{\sigma_i}^2}\right],
\label{clik}
\end{equation}
where $\vec{x}_i$ is the location of the $i$th event, ${\sigma_i}^2$ is
its Gaussian variance, $\Gamma^{-2}\equiv\sum_{i=1}^n{\sigma_i}^{-2}$
and $\vec{z}\equiv\Gamma^2\sum_{i=1}^n\vec{x}_i/{\sigma_i}^2$.

The usefulness of $\mathcal{L}$ for clustering studies is apparent from
its form.  It is a product of two competing factors:   The first,
rational factor increases with increasing $n$ (recall that the $\sigma$
are generally much less than 1), and so rewards larger putative
clusters.  At the same time, the second, exponential factor penalizes
clusters that are too dispersed.  This tension makes $\mathcal{L}$ an
excellent clustering statistic.  Sadly, its distribution function is
not known, and must be obtained by Monte Carlo.  

We have performed the simulations, creating BATSE catalogs with
uniformly distributed, non-repeating bursts whose location errors are
randomly drawn from the BATSE location error distribution.  In the case
of $n$ events in a cluster, we record the number of times that $n$ or
more events occur within 1.8 days or less and have a value of
$\mathcal{L}$ exceeding that of the data. In the case of four events,
the probability that we obtain is $3.1\times 10^{-5}$, while in the
three-event case the probability is $1.6\times 10^{-2}$.

\section*{Conclusions}

\begin{table}
\caption{Summary of Statistical Significances}
\label{table1}
\begin{tabular}{ldd}
& minimum Circle Probability & Cluster Likelihood Probability \\
\tableline
Three Events & 1.2$\times 10^{-1}$ & 1.6$\times 10^{-2}$ \\
Four Events  & 3.3$\times 10^{-4}$ & 3.1$\times 10^{-5}$ \\
\end{tabular}
\end{table}
Our results are summarized in Table \ref{table1}.  We find that the
evidence in favor of repetition of burst sources depends very strongly
upon the assumed number of events.  Assuming four events, the evidence
is quite strong.  In the case of three events, the significance is much
weaker, but in this case one must accept that one of the events was the
longest gamma-ray burst ever observed by BATSE.

We acknowledge support from NASA grants NAGW-4690, NAG 5-1454, and NAG
5-4406.


\begin{references}

\bibitem{fishman94} Fishman,~G.~J., et al. 1994, {\it ApJ Suppl.} {\bf
92},  229
%
\bibitem{lbc95} Lamb,~D.~Q., Bulik,~T., and Coppi,~P.~S. 1996, in {\it
High-Velocity Neutron Stars and Gamma-Ray Bursts}:  AIP Conference
Proceedings 366, R.~E. Rothschild and R.~E.~Lingenfelter, eds., p. 219
%
\bibitem{lamb96} Lamb,~D.~Q. 1996, in {\it Gamma-Ray Bursts}:  AIP
Conference Proceedings 384, C.~Kouveliotou, M.~F.~Briggs, and
G.~J.~Fishman, eds., p. 281
%
\bibitem{vc97} Connaughton,~V. {\it et al.} 1997, in {\it Proceedings of
the 18th Texas Symposium on Relativistic Astrophysics}, A.~Olinto,
J.~Frieman, and D.~Schramm, eds., in press
%
\bibitem{gl96} Graziani,~C., and Lamb,~D.~Q. 1996, in {\it Gamma-Ray
Bursts}:  AIP Conference Proceedings 384, C.~Kouveliotou, M.~F.~Briggs,
and G.~J.~Fishman, eds., p. 382
%
\bibitem{gl96b} Graziani,~C., and Lamb,~D.~Q. 1996, in {\it
High-Velocity Neutron Stars and Gamma-Ray Bursts}:  AIP Conference
Proceedings 366, R.~E. Rothschild and R.~E.~Lingenfelter, eds., p. 196

\end{references}
\end{document}